\documentstyle[pre,multicol,aps,epsfig]{revtex}

\begin{document}

\draft

\tighten

\title{Overdamped sine-Gordon kink in a thermal bath}

\author{Niurka R.\ Quintero$^{*}$ and Angel S\'anchez$^{\dag}$} 

\address{Grupo Interdisciplinar de Sistemas
Complicados (GISC), Departamento de Matem\'aticas, 
Universidad Carlos III de Madrid,\\
Edificio Sabatini,
Avenida de la Universidad 30, E-28911 Legan\'{e}s, Madrid, Spain}

\author{Franz G.\ Mertens$^{\ddag}$}

\address{Physikalisches Institut, Universit\"at Bayreuth, 
D-95440 Bayreuth, Germany}

\date{\today}

\maketitle

\begin{abstract} 
We study the sine-Gordon kink diffusion at finite temperature 
in the overdamped limit. By means of a general perturbative approach,
we calculate the first- and second-order (in temperature)
contributions to the diffusion coefficient. We compare our analytical
predictions with numerical simulations. The  
good agreement allows us to conclude that, up to temperatures where
kink-antikink nucleation processes cannot be neglected, a diffusion 
constant linear and {\it{quadratic}} in temperature gives a very accurate description of
the diffusive motion of the kink. The quadratic temperature 
dependence is shown to stem from the interaction with the phonons. 
In addition, we calculate and compute 
the average value $\langle \phi(x,t) \rangle$ of the wave function as a function of time and show that 
its width grows with $\sqrt{t}$. We discuss the interpretation
of this finding and show that it arises from the dispersion of the kink 
center positions of individual realizations which all keep their width. 
\end{abstract}

\pacs{PACS: 
{03.40.Kf}, 
{05.40.+j}, 
{74.50.+r}, 
{85.25.Cp}
} 

\widetext

\section{Introduction}
\label{intro}

There is no longer any controversy about the physical relevance of noise
effects in spatially extended, nonlinear systems \cite{Bishop,GSbook}: 
Indeed, the pervasive, joint role of nonlinearity and (static or dynamic)
disorder has already been recognized in biophysics, electronics, optics,
fluids, condensed matter, computational physics, etc. In most of these 
fields, nonlinear phenomena involve nonlinear coherent excitations, such
as solitons or solitary waves, which play a key part in the corresponding
system dynamics. It is because of this nowadays well established fact that
much effort has been devoted to understand how stochastic perturbations 
affect solitons, mostly during the decade of the 80's (see \cite{Bass,%
yo1,KVbook} for reviews). In fact, early numerical simulations 
\cite{early} already revealed that $\phi^4$ solitary waves underwent 
Brownian-like motion in the presence of additive white noise, i.e., of thermal 
fluctuations. Subsequent works focused on the study of soliton diffusion,
since it may be crucial in a number of problems, such as photoexcitation 
dynamics, photoconductivity of conducting polymers, or transport by phase
solitons in charge-density-wave systems, to name a few \cite{aplic}.

Among the different soliton-bearing nonlinear models which have been 
studied in the above context, one which has received a great deal of 
attention is the sG equation. The interest in this model is both
theoretical, as it displays the main features of more realistic and 
complicated cases while remaining analytically tractable, and applied,
as it very approximately describes the dynamics of many physically 
relevant systems, such as one-dimensional (1D) magnets \cite{mikeska} or 
long Josephson junctions \cite{barone}, for instance. Soliton diffusion 
governed by the sG (and other nonlinear Klein-Gordon equations)
has been studied along two main, different lines which are discussed and
compared, e.g., in Ref. \cite{ivanov}. 
The first one consists
of considering extended excitations of the system (phonons) in equilibrium
with both a single sG soliton and a heat bath at temperature $T$.  
This approach leads to two distinct diffusion regimes: anomalous diffusion,
characterized by a diffusion constant proportional to $T^2$, and viscous 
diffusion, when the appearance of a dynamical damping coefficient yields
a diffusion constant proportional to $T^{-1}$. We will not follow this 
approach here; the interested reader is referred to the detailed review 
by Y.\ Wada \cite{wada}. 
The second manner is {\em \`a
la} Langevin, i.e., introducing the influence of an external thermal bath 
by means of local fluctuations of the string and a local damping force 
related to that by an appropriate fluctuation-dissipation relationship. 
The corresponding equation of motion is then

\begin{equation}
\phi_{tt} - \phi_{xx} + \sin(\phi) = -\alpha \phi_{t} + \eta(x,t),
\label{ecua1}
\end{equation}
with
\begin{mathletters}
\begin{eqnarray}
\langle\eta(x,t)\rangle &=& 0,  \\
\langle\eta(x,t)\eta(x',t')\rangle &=& D \delta(x-x') \delta(t-t'), 
\end{eqnarray} \noindent
\label{ecua3} 
\end{mathletters}

\noindent
where the diffusion coefficient 
$D=2 \alpha k_{b} T$, $k_{b}$ being the Boltzmann constant,
and $-\alpha \phi_{t}$ being the damping term with a dissipation 
coefficient $\alpha$. This equation has been considered a number of times
in the literature (see, e.g., \cite{Bass} and references therein; see also 
\cite{cast} for related experimental work). 

In this work, we focus on the Langevin version of the problem, with the 
aim of improving the analytical results obtained in the aforementioned 
works as well as of verifying them by numerical simulations specifically 
planned to that end. Furthermore, we concern ourselves with the overdamped
limit of the sG equation, which reads 
\begin{equation}
\alpha \phi_{t} - \phi_{xx} + \sin(\phi) = \epsilon\eta(x,t,\phi_,...). 
\label{ecua2}
\end{equation} 
Note that we have introduced a factor $\epsilon$ in front of the noise 
term for convenience in the analytical calculations in section II.
This equation (without noise, $\epsilon=0$) was already considered by 
Eilenberger in 
\cite{bre}, as the limit of the sG equation (\ref{ecua1})
in the case when the dissipation effect 
is strong enough in Eq.\ (\ref{ecua1}) and there is an input of energy 
into the system (see, e.g., 
\cite{km,bennet} and references therein). 
On the other hand, Eq.\ (\ref{ecua2}), with additive noise as in 
(\ref{ecua3}), is interesting in itself:
For example, it has been proposed as a model for crystal growth (see
\cite{ks,ancai,rangel} and references therein). Equation (\ref{ecua2}) 
has also been studied to analyze the kink contribution to transport 
properties when the system is driven and thermally activated  
\cite{ks,butland1,butland2,kaup}. In particular, the work of Kaup \cite{kaup} is the most
closely related to the present one, as it presents a singular perturbation
theory to compute the first-order (in $T$) correction to the kink 
mobility as well as the 
change of its shape. However, to our knowledge the free diffusion problem 
for the overdamped sG equation 
has not been adressed in the literature to date and, therefore, we believe that
our results will be interesting by themselves. On the other hand, 
we also hope that what we 
learn in this case can be used towards obtaining a more complete, accurate
picture of the full sG problem; we will discuss this question 
in the conclusions. 

The outline of the paper is as follows:
In section II, 
using a general perturbative method \cite{bre} which we recall in detail,
we calculate the correlation functions of the position and the velocity of 
the kink center up to second-order in 
$k_{b} T$, as well as the diffusion coefficient and the mean value 
$\langle\phi(x,t)\rangle$ for fixed $t$. 
In section III we numerically 
integrate the stochastic partial differential equation 
(\ref{ecua2}), with noise given by Eq.\ (\ref{ecua3}),   
using the Heun scheme \cite{maxra} and    
compute the time correlation function of 
the position of the kink center and the diffusion coefficient. We  
compare these results with the theoretical ones 
obtained in section II and find an excellent agreement. Finally, 
in section IV we discuss our results, summarize our main conclusions,
and sketch lines for future research.  

\section{A general perturbative approach}

Following the {\em Ansatz} proposed in \cite{bre,raj}, we assume that the 
solution of Eq.\ (\ref{ecua2}) can be expanded as 
\begin{equation}
\phi(x,t)=\phi_{0}[x-X(t)] + \int_{-\infty}^{+\infty}dk\,A_{k}(t)
f_{k}[x-X(t)],
\label{ecua4}
\end{equation}
where $f_{k}[x-X(t)]$ are the eigenfunctions 
of the linearized version of Eq.\ (\ref{ecua2}) [with $\epsilon=0$], which  
along with $\displaystyle 
f_{T} [x-X(t)] = \frac{\partial{\phi_{0}}}{\partial{x}}[x-X(t)]$, 
form a complete set of orthogonal eigenfunctions (see Appendix I).  The first 
term in the expansion (\ref{ecua4}) represents the translational mode related to 
the position of the kink center $X(t)$, whereas the second one 
characterizes the phonon modes (linear excitations around a kink) 
of the system. We will focus on the kink center motion as described by 
$X(t)$, as it is well established that such a particle-like picture is 
very generally enough to describe the behavior of the kink as a whole
($X$ playing the r\^ole of a collective coordinate; see, e.g., \cite{siam}
for a review). 

In order to calculate the dynamics of the kink center, we begin by 
inserting (\ref{ecua4}) in (\ref{ecua2}), 
and projecting on the orthogonal 
basis [see Appendix I, relationships (\ref{ap6})] we obtain 
a system of differential equations for the unknown 
functions $X(t)$ and $A_{k}(t)$:

\begin{eqnarray}
\dot{X}(t) & = & - 
\frac{1}{8} \dot{X}(t) \int_{-\infty}^{+\infty}dk A_{k}(t) I_{1}(k) -  
\frac{1}{16 \alpha} \int_{-\infty}^{+\infty} dk \int_{-\infty}^{+\infty}dk' A_{k}(t) A_{k'}(t) R_{3}(k,k') + \nonumber \\ 
& + & \frac{\sqrt{D}}{8 \alpha} \int_{-\infty}^{+\infty} f_{T} [x - X(t)] \,\ \eta(x,t) \,\ dx - \nonumber \\
 & -& 
\frac{1}{48 \alpha}
\int_{-\infty}^{+\infty} dk \int_{-\infty}^{+\infty} dk_{1} \int_{-\infty}^{+\infty} dk_{2} A_{k}(t) A_{k_{1}}(t) A_{k_{2}}(t) 
R_{6}(k,k_{1},k_{2}),\label{ecua5}\\
\frac{\partial{A_{k}}}{\partial{t}} + \frac{\omega_{k}^{2}}{\alpha} A_{k}(t) & = & 
\dot{X}(t) \int_{-\infty}^{+\infty} dk A_{k}(t) I_{3}(k,k') + \frac{1}{2 \alpha} \int_{-\infty}^{+\infty} dk 
\int_{-\infty}^{+\infty} dk'
A_{k}(t) A_{k'}(t) R_{4}(k,k') - \nonumber \\ 
& - & \frac{\sqrt{D}}{\alpha} \int_{-\infty}^{+\infty} 
f_{k'}^{*} [x - X(t)] \,\ \eta(x,t) \,\ dx + \nonumber \\
& + & 
\frac{1}{6 \alpha} 
\int_{-\infty}^{+\infty} dk \int_{-\infty}^{+\infty} dk_{1} \int_{-\infty}^{+\infty} dk_{2} A_{k}(t) A_{k_{1}}(t) 
A_{k_{2}}(t) R_{7}(k,k_{1},k_{2}),
\label{ecua6}
\end{eqnarray}
where 
\begin{eqnarray}
I_{1}(k) & = & \int_{-\infty}^{+\infty} 
\frac{\partial f_{k}}{\partial \theta} f_{T}(\theta) d\theta = 
\frac{i \pi \omega_{k}}{\sqrt{2 \pi} \,\ \cosh\Big(\displaystyle{\frac{\pi k}{2}}\Big)}, \nonumber \\
R_{3}(k,k') & = & \int_{-\infty}^{+\infty} f_{T}(\theta) 
\frac{\partial f_{T}}{\partial \theta} f_{k}(\theta) 
f_{k'}^{*}(\theta) d\theta =  
-\frac{i (\omega_{k}^{2} - \omega_{k'}^{2})^{2}}{4 \omega_{k} \omega_{k'} 
\,\ \sinh\Big(\displaystyle{\frac{\pi \Delta k}{2}}\Big)},\,\ \Delta k=k'-k, \nonumber \\
I_{3}(k,k') & = & \int_{-\infty}^{+\infty} 
\frac{\partial f_{k}}{\partial \theta} f_{k'}^{*}(\theta) d\theta, \nonumber \\
R_{4}(k,k') & = & \int_{-\infty}^{+\infty} [f_{k'}^{*}(\theta)]^{2}   
\frac{\partial f_{T}}{\partial \theta} f_{k}(\theta) 
 d\theta, \,\ R_{4}(k,k) = \frac{3 i \omega_{k}}{8 \sqrt{2 \pi} \,\ 
\cosh\Big(\displaystyle{\frac{\pi k}{2}}\Big)}, \nonumber \\
R_{6}(k,k_{1},k_{2}) & = & \int_{-\infty}^{+\infty} 
\frac{\partial^{2} f_{T}}{\partial \theta^{2}} f_{k}(\theta) f_{k_1}^{*}(\theta) 
f_{k_2}(\theta) d\theta, \,\ \nonumber \\
R_{7}(k,k_{1},k_{2}) & = & \int_{-\infty}^{+\infty} \cos(\phi_{0}) 
f_{k'}^{*}(\theta) f_{k}(\theta) f_{k_1}^{*}(\theta) f_{k_2}(\theta)
 d\theta.
\label{ecua7}
\end{eqnarray}

We now recall that, if we set $\epsilon = 0$ in (\ref{ecua2}), 
the static kink is an exact solution; hence, in what follows we will 
consider $\epsilon$ as a small perturbative parameter, and 
expand $A_{k}(t)$ and $X(t)$ in powers of $\epsilon$. 
By substituting the series $A_{k}(t) = \sum_{n=1}^{\infty} \epsilon^{n} A_{k}^{n}(t)$ and $X(t) = \sum_{n=1}^{\infty} \epsilon^{n} X_{n}(t)$ in 
(\ref{ecua5}) and (\ref{ecua6}) we find a set of linear 
equations for the coefficients of these series. We only write down here
the systems of equations up to order $\epsilon^{3}$, leaving out the 
cumbersome (albeit straightforward) equation for $A_{k}^{3}(t)$:

\underline{$O(\epsilon)$}
\begin{mathletters}
\begin{eqnarray}
\dot{X}_{1}(t) & = & \epsilon_{1}(t), \,\ \langle \epsilon_{1}(t)\rangle =0, \,\ 
\langle \epsilon_1(t) \epsilon_1(t')\rangle  = \frac{D}{8 \alpha^{2}} \delta(t-t'), \label{ecua8}\\
\frac{\partial {A_{k}^{1}}}{\partial t}(t) + \frac{\omega_{k}^{2}}{\alpha} 
A_{k}^{1}(t) & = & \frac{\epsilon_{k}(t)}{\alpha}, \,\ \langle\epsilon_{k}(t)\rangle =0, \,\ 
\langle\epsilon_{k}(t) \epsilon_{k'}(t')\rangle  = \frac{D}{\alpha^{2}} \delta(t-t') 
\delta(k-k');
\label{ecua9}
\end{eqnarray} 
\end{mathletters}

\underline{$O(\epsilon^2)$}
\begin{mathletters}
\begin{eqnarray}
\dot{X}_{2}(t) & = & -\frac{\dot{X}_{1}(t)}{8} \int _{-\infty}^{+\infty} dk A_{k}^{1}(t) I_{1}(k) - \nonumber \\
& - & \frac{1}{16 \alpha} \int_{-\infty}^{+\infty} dk \int_{-\infty}^{+\infty} dk' 
A_{k}^{1}(t) A_{k'}^{1}(t) R_{3}(k,k'), \label{ecua10}\\
\frac{\partial {A_{k}^{2}}}{\partial t}(t) + \frac{\omega_{k}^{2}}{\alpha} 
A_{k}^{2}(t) & = & \dot{X}_{1}(t) \int_{-\infty}^{+\infty} dk A_{k}^{1}(t) I_{3}(k,k) + \nonumber \\ 
& + & \frac{1}{2 \alpha} \int_{-\infty}^{+\infty} dk \int_{-\infty}^{+\infty} dk' 
A_{k}^{1}(t) A_{k'}^{1}(t) R_{4}(k,k');
\label{ecua11}
\end{eqnarray} 
\end{mathletters}

\underline{$O(\epsilon^{3})$}
\begin{eqnarray}
\dot{X}_{3}(t) & = & 
-\frac{\dot{X}_{1}(t)}{8} \int_{-\infty}^{+\infty} dk  A_{k}^{2}(t) I_{1}(k) - 
\frac{\dot{X}_{2}(t)}{8} \int_{-\infty}^{+\infty} dk A_{k}^{1}(t) I_{1}(k) - \nonumber \\
& - & \frac{1}{16 \alpha} \int_{-\infty}^{+\infty} dk \int_{-\infty}^{+\infty} dk' 
A_{k}^{2}(t) A_{k'}^{1}(t) R_{3}(k,k') - \nonumber \\
& - & \frac{1}{16 \alpha} \int_{-\infty}^{+\infty} dk \int_{-\infty}^{+\infty} dk' 
A_{k}^{1}(t) A_{k'}^{2}(t) R_{3}(k,k') - \nonumber \\ 
& - &\frac{1}{48 \alpha} \int_{-\infty}^{+\infty} dk \int_{-\infty}^{+\infty} dk_{1} \int_{-\infty}^{+\infty} dk_{2} A_{k}^{1}(t) 
A_{k_{1}}^{1}(t) A_{k_{2}}^{1}(t) R_{6}(k,k_{1},k_{2}). \label{ecua12} 
\end{eqnarray} 

We now proceed with the first-order equations. 
The solutions of (\ref{ecua8}) and (\ref{ecua9}) can be written as
\begin{eqnarray}
X_{1}(t) & = & \int_{0}^{t} \epsilon_{1}(\tau) d \tau, \quad 
A_{k}^{1}(t) = \exp(-\frac{\omega_{k}^{2} \,\ t}{\alpha}) 
\int_{0}^{t} \exp(\frac{\omega_{k}^{2} \tau}{\alpha}) \epsilon_{k}(\tau) d \tau, 
\label{ecua13}
\end{eqnarray}
respectively. From these relations we can immediately 
compute averages over the quantities of interest, such as  
\begin{eqnarray}
\langle X_{1}(t)\rangle  & = & 0, \,\ \langle X_{1}(t) X_{1}(t')\rangle  = 
\frac{D}{8 \alpha^{2}} M, \label{ecua14}\\ 
\langle \dot{X}_{1}(t)\rangle  & = & 0, \,\ \langle \dot{X}_{1}(t) \dot{X}_{1}(t')\rangle  = 
\frac{D}{8 \alpha^{2}} \delta(t-t'),\label{ecua15}\\ 
\langle A_{k}^{1}(t)\rangle  & = & 0, \quad \langle A_{k}^{1}(t) A_{k}^{1}(t')\rangle  = 
\frac{D}{2 \alpha \omega_{k}^{2}} 
\Big[\exp\Big(-\frac{\omega_{k}^{2} |t'-t|}{\alpha}\Big) - 
\exp\Big(-\frac{\omega_{k}^{2} (t+t')}{\alpha}\Big)\Big],  
\label{ecua16}
\end{eqnarray} 
where $M=min(t,t')$.
For the next orders, the calculations are more involved but not difficult. 
After some tedious algebra, 
from Eqs.\ (\ref{ecua10})-(\ref{ecua12}) we find 
the average values of the position and velocity of the kink center  
\begin{eqnarray}
\langle X_{2}(t)\rangle  & = & 0, \,\langle \dot{X}_{2}(t)\rangle  = 0, \label{ecua17}\\
\langle X_{3}(t)\rangle  & = & 0, \,\langle \dot{X}_{3}(t)\rangle  = 0; 
\label{ecua18}
\end{eqnarray}
whereas it can be shown that, for large enough times,
\begin{mathletters}
\begin{eqnarray}
\langle |A_{k}^{2}(t)|\rangle &\sim &
\frac{3 \sigma k_{b} T}{16 \sqrt{2 \pi} \omega_{k}^{2}},\\
\sigma &=& \int_{-\infty}^{+\infty} \frac{dk}{\omega_{k} 
\cosh\Big(\displaystyle{\frac{\pi k}{2}}\Big)} \approx 1.62386.     
\end{eqnarray}
\end{mathletters}
The corresponding correlation functions for $X_{2}(t)$ and $\dot{X}_{2}(t)$ are 
\begin{eqnarray}
\langle X_{2}(t) X_{2}(t')\rangle  & = & \frac{D^{2} M}{512 \alpha^{3}} + 
\frac{D^{2} \pi}{4096 \alpha^{2}} \int_{-\infty}^{+\infty} 
\frac{\Big[\exp\Big(-2 \omega_{k}^{2} M/\alpha \Big)-1 \Big] dk}{\omega_{k}^{2} 
\cosh^{2}\displaystyle{\Big(\frac{\pi k}{2}\Big)}}, \label{ecua19}\\
\langle \dot{X}_{2}(t) \dot{X}_{2}(t')\rangle  & = & \langle \dot{X}_{1}(t) \dot{X}_{1}(t')\rangle  \times \nonumber \\
& \times &
\frac{D \pi}{256 \alpha} \int_{-\infty}^{+\infty} 
\frac{\exp\Big(- \omega_{k}^{2} |t'-t|/\alpha \Big) - 
\exp\Big(- \omega_{k}^{2} (t'+t)/\alpha \Big) dk}
{\cosh^{2}\displaystyle{\Big(\frac{\pi k}{2}\Big)}}. 
\label{ecua20}
\end{eqnarray} 

Notice that the cross correlation function of $X_{1}(t)$ and 
$X_{3}(t)$ is of the same order as $\langle X_{2}(t) X_{2}(t')\rangle $, 
and also 
that $\langle X_{1}(t) X_{2}(t')\rangle =0$. So, from Eqs.\ (\ref{ecua8}) and (\ref{ecua12}) 
we have 

\begin{equation}
\begin{array}{l}
\displaystyle {\langle X_{3}(t) X_{1}(t')\rangle  = \langle X_{2}(t) X_{2}(t')\rangle  - \frac{D^{2}}{256 \alpha^{3}} 
\int_{-\infty}^{+\infty} dk \,\ I_{1}(k)} \times \\
\\
\displaystyle {\times \Big\{ \Big( 
\int_{-\infty}^{+\infty} dm \frac{R_{4}(m,m)}{\omega_{m}^{2}} \Big) \,\
\Big[\frac{M}{\omega_{k}^{2}} + 
\frac{\alpha (\exp(-\omega_{k}^{2} M/\alpha)-1)}{\omega_{k}^{4}} \Big]} - \\
\\
\displaystyle {- \int_{-\infty}^{+\infty} dn 
\frac{R_{4}(n,n)}{\omega_{n}^{2}} \,\
\frac{\alpha}{2 \omega_{n}^{2}-\omega_{k}^{2}} 
\Big[\frac{(\exp(-2 \omega_{n}^{2} M/\alpha)-1)}{2 \omega_{n}^{2}} - 
\frac{(\exp(-\omega_{k}^{2} M/\alpha)-1)}{\omega_{k}^{2}}\Big] \Big\}}, 
\label{ecua21}
\end{array} 
\end{equation}

\begin{equation}
\begin{array}{l}
\displaystyle {\langle \dot{X}_{3}(t) \dot{X}_{1}(t')\rangle  = - \frac{1}{8} 
\langle \dot{X}_{1}(t) \dot{X}_{1}(t')\rangle  \times } \,\ \nonumber \\
\displaystyle{\times \int_{-\infty}^{+\infty} dk\Big \{ 
\langle A_{k}^{2}(t)\rangle  I_{1}(k) - \frac{1}{8} \langle [A_{k}^{1}(t)]^{2}\rangle  |I_{1}(k)|^{2} 
\Big \}}.
\label{ecua22}
\end{array}
\end{equation}
Finally, from (\ref{ecua15}), (\ref{ecua20}) and (\ref{ecua22}) we obtain 
the final result, namely
that for large $t$ [i.e., taking the limit $t\to\infty$ in Eqs.\ 
(\ref{ecua20}) and (\ref{ecua22}) in all terms except those related to 
$X_1(t)$] the correlation function 
$\langle \dot{X}(t) \dot{X}(t')\rangle $ is given up to 
order $\epsilon^{4}$ by   
\begin{eqnarray}
\langle \dot{X}(t) \dot{X}(t')\rangle  & = & 
\epsilon^{2} \langle \dot{X}_{1}(t) \dot{X}_{1}(t')\rangle  + \nonumber \\
& + & 
\epsilon^{4} (\langle \dot{X}_{2}(t) \dot{X}_{2}(t')\rangle  + 
\langle \dot{X}_{1}(t) \dot{X}_{3}(t')\rangle  + \langle \dot{X}_{3}(t) \dot{X}_{1}(t')\rangle )+... \nonumber \\
& = &  
\frac{\epsilon^{2}}{8} 
\langle \dot{X}_{1}(t) \dot{X}_{1}(t')\rangle  \Big\{ 1+ \epsilon^{2} \Big(\frac{3}{32} + \frac{3}{128} 
\sigma^{2}\Big) k_{b} T \Big \} + o(\epsilon^{4}).  
\label{ecua23}
\end{eqnarray}

We now return to our original equation notation: We set
$\epsilon$ equal to one and consider $\sqrt{k_{b} T}$ as the small parameter. 
When $t$ goes to infinity and imposing $\epsilon=1$, from  
Eqs.\ (\ref{ecua14}), (\ref{ecua19}) and 
(\ref{ecua21}) we find  

\begin{eqnarray}
\langle [X(t)]^{2}\rangle  & = & 
\frac{k_{b} T}{4 \alpha} t \Big\{ 1+ \Big(\frac{3}{32} + \frac{3}{128} 
\sigma^{2}\Big) k_{b} T \Big \}. 
\label{corrx}
\end{eqnarray}
Note that the slope of this function is the kink diffusion coefficient, so if one 
takes into account 
the second-order correction one obtains that the diffusion coefficient is a 
quadratic function of the temperature. We postpone our comments to Sec. IV, where a comparison with the previously available 
results will be made. 

To complete this work, we can calculate in a very simple way 
the average value of the wave 
function $\phi(x,t)$ in first order:  
{}From Eq.\ (4) we have that 

\begin{equation}
\langle \phi(x,t)\rangle =\langle \phi_{0}[x-\epsilon X_1(t)]\rangle + O(\epsilon^{2}).
\label{avphi}
\end{equation}
In this last relation we have taken into account that 
$\langle A_{k}(t)\rangle =\epsilon \langle A_{k}^{1}(t)\rangle  + 
O(\epsilon^{2})$ and $\langle A_{k}^{1}(t)\rangle =0$ [see Eq.\ (\ref{ecua16})]. 

If we solve the corresponding Fokker-Planck equation for $X_{1}$ [see Eq.\ (\ref{ecua8})], 
we obtain that the probability distribution function for $X_{1}$  is a Gaussian function given by 

\begin{equation}
p(X_{1})=\sqrt{\frac{4 \alpha^{2}}{\pi t D}} 
\exp\Big(-\frac{4 \alpha^{2} X_{1}^{2} }{D t} \Big).
\label{pdf}
\end{equation}

So, one can define the average value $\langle \phi(x,t)\rangle $ as 

\begin{equation}
\langle \phi(x,t)\rangle = \int_{-\infty}^{+\infty} dX_{1} p(X_{1}) \phi_{0}[x-\epsilon X_{1}(t)].
\label{inte}
\end{equation}
Unfortunately we have not found the analytical expression for this integral.
But   
we have calculated it numerically, and below we will compare it to the 
simulations for the full partial differential equation. 

\section{Numerical simulations}

For our numerical simulations of the partial differential equation 
(\ref{ecua2}), we have used 
the Heun method \cite{maxra}, whose stochastic properties are well 
known and suitable for comparison to our theoretical predictions. We 
numerically integrate Eqs.\ (\ref{ecua2}), with white noise (\ref{ecua3}),  
starting from an unperturbed kink 
at rest and taking the average values over 1000 realizations. 
The other parameters are 
$\alpha=1$, $\Delta x = 0.05$,
and $\Delta t = 0.001$. In the evaluation of the simulations, we have defined the 
center of the kink as follows: We first find all the lattice points $i$ such that 
$\phi_{i} \le \pi$ and $\phi_{i+1} \ge \pi$ or vice versa. 
We then interpolate to obtain the points $x_{i}$ where the field $\phi$ crosses $\pi$. In case 
that, due to the noise-induced deformation of the kink, there is more than one 
such $x_{i}$, we average them to finally obtain the numerical kink center 
position, $x_{c}$. As discussed below, this introduces some error, but
other alternatives we tested (such as the center of mass, for instance)
gave results which did not really represent the kink location, and 
moreover its calculation from numerics is much less accurate.    
Once the center is obtained, we computed 
also its variance $\langle [X(t)]^{2}\rangle $. 

\begin{figure}
\begin{center}
\hspace{-2.4cm}
\epsfig{file=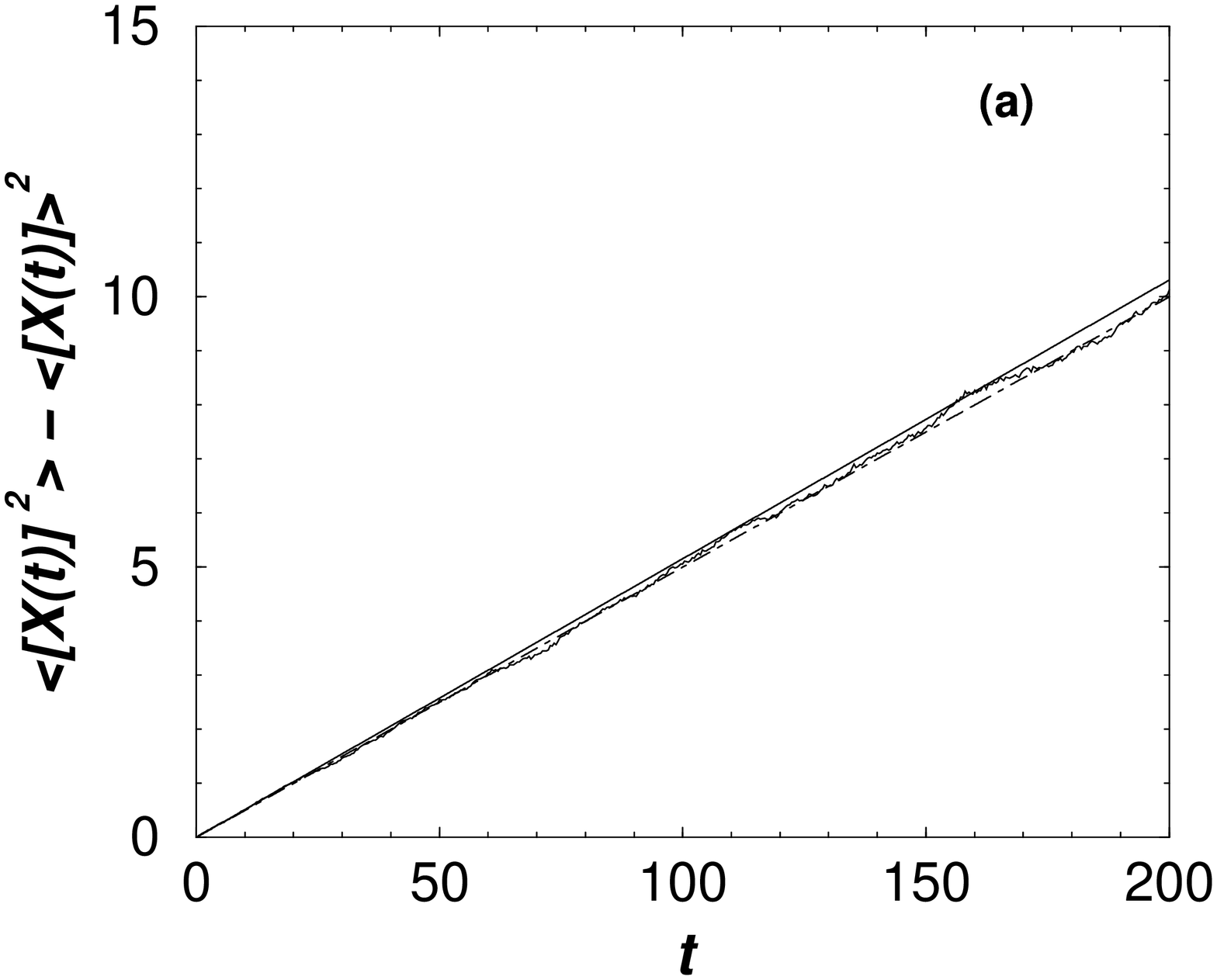, width=2.7in, angle=-90}
\epsfig{file=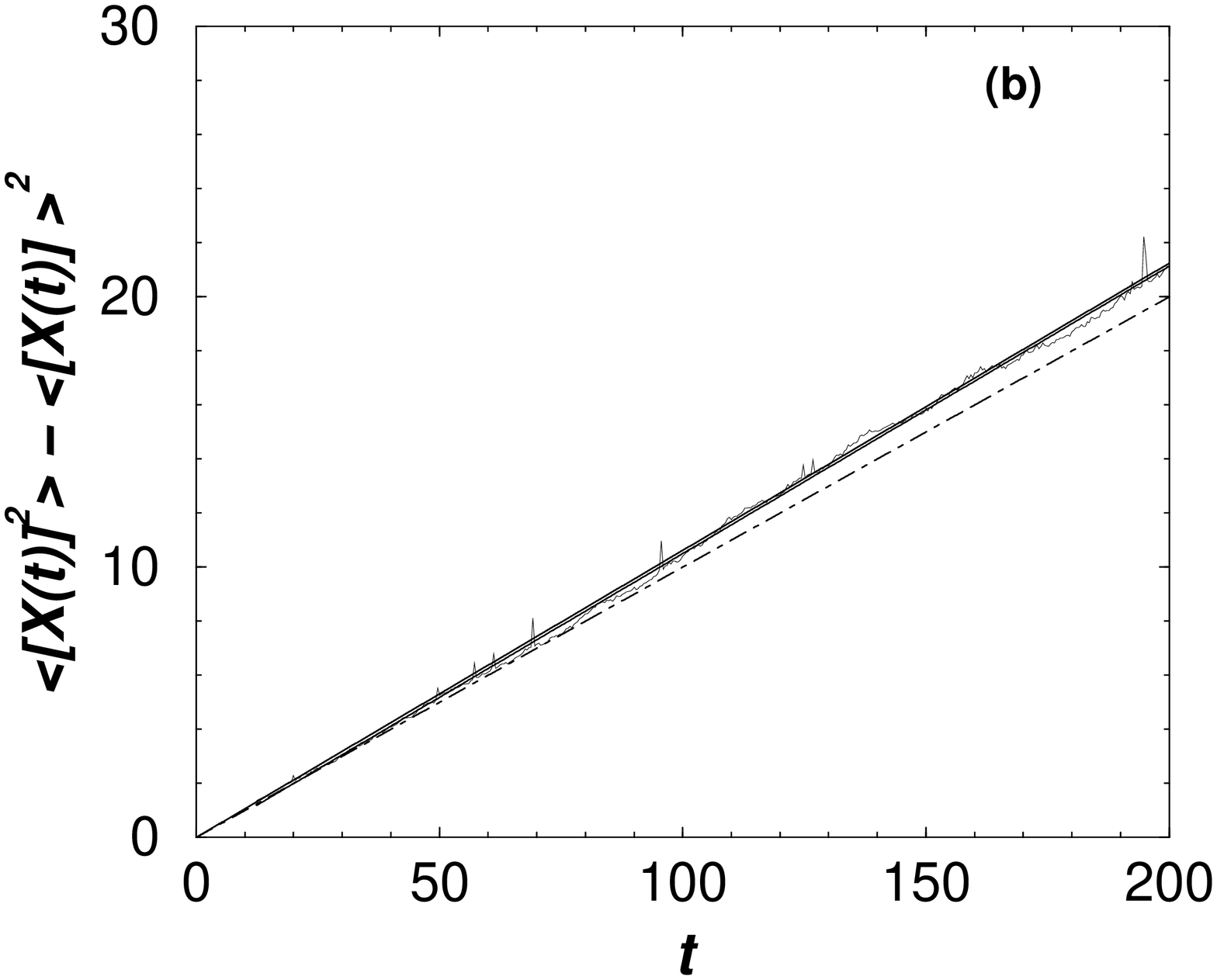, width=2.7in, angle=-90}\\
\hspace{-2.4cm}
\epsfig{file=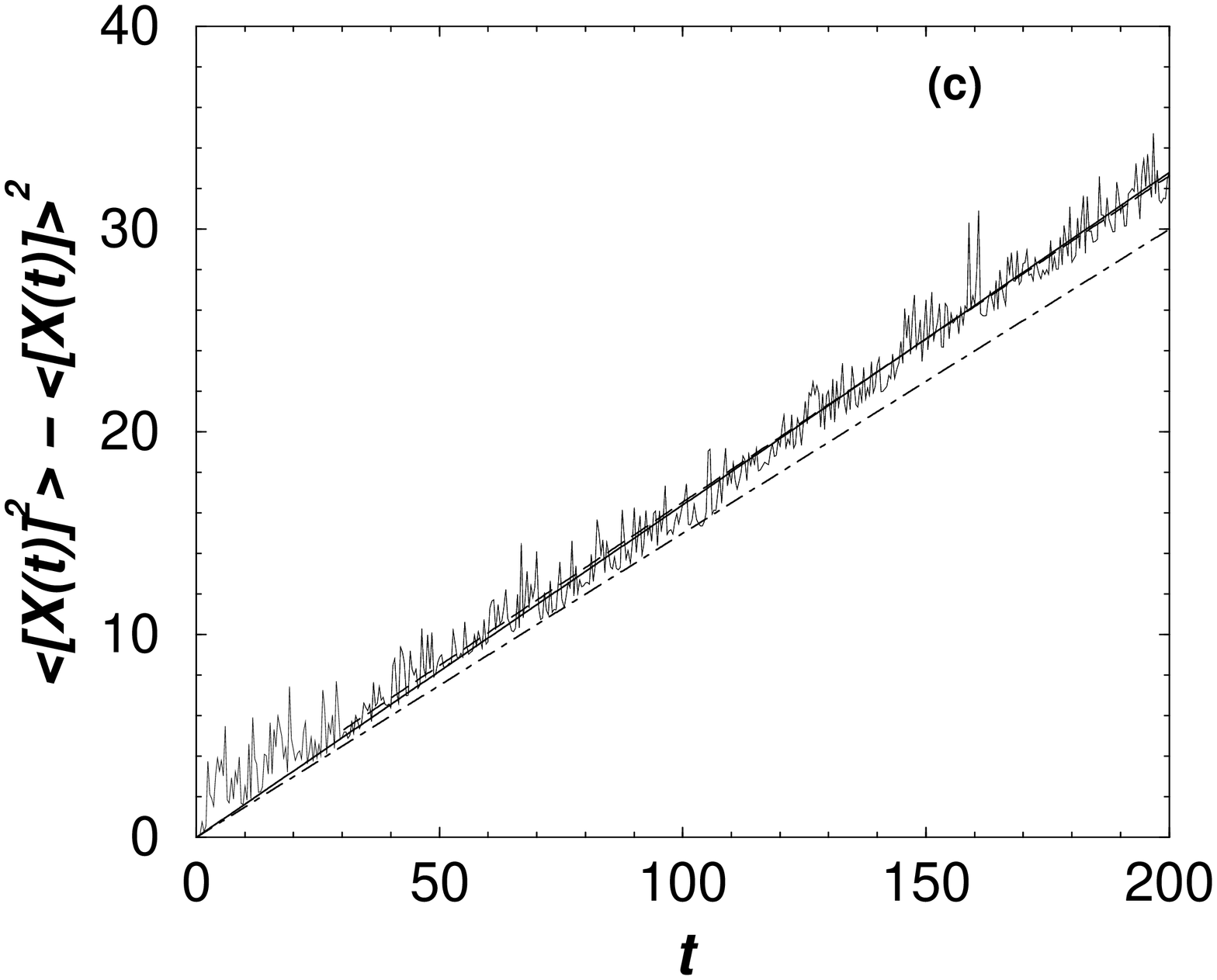, width=2.7in, angle=-90}
\epsfig{file=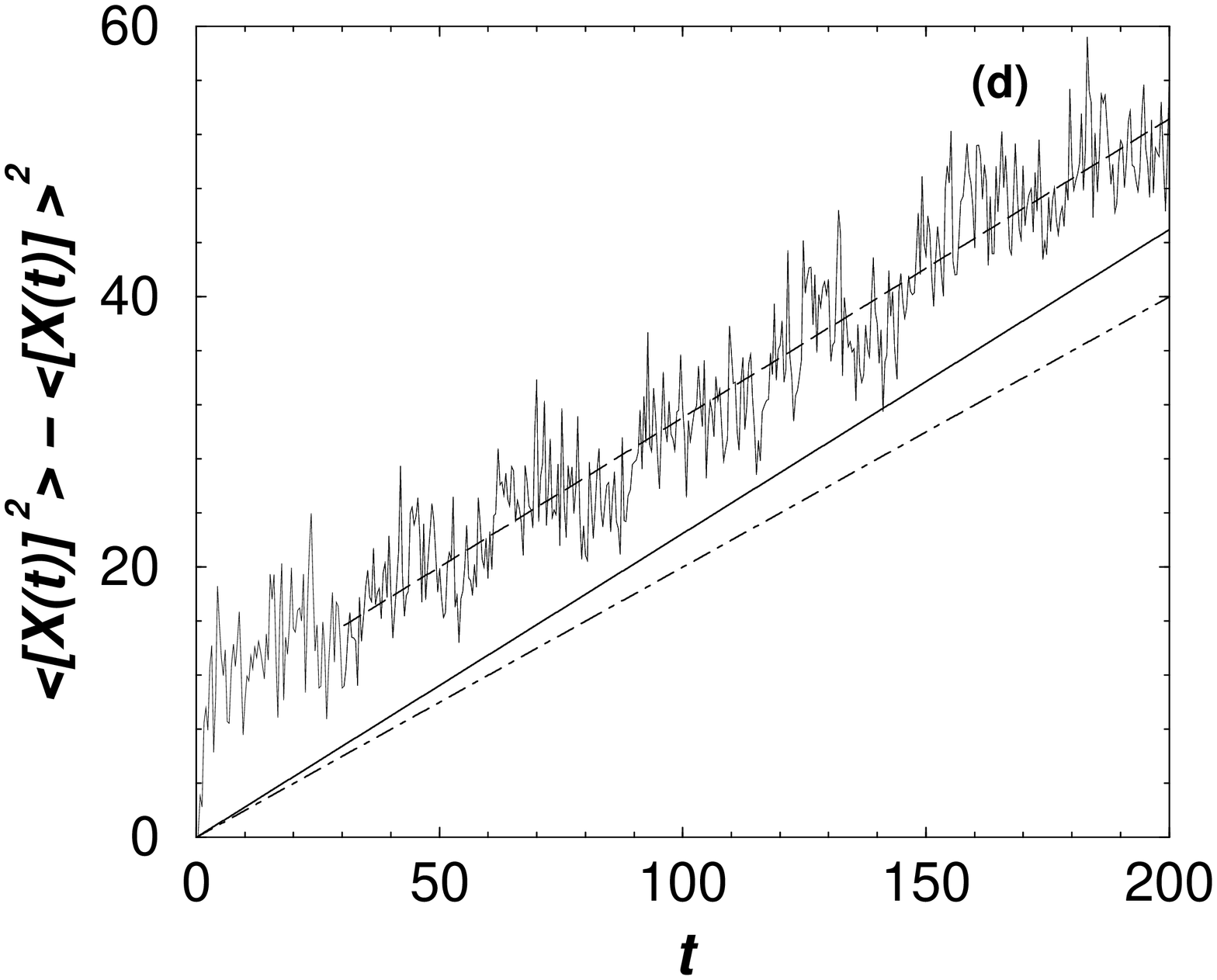, width=2.7in, angle=-90}\\
\ \\[15mm]
\caption[]{Simulations with initial condition given by 
a static kink initially located at 
$X(0)=0$, and subject to a thermal bath. 
As a continuous  
(but wiggly) line, we have plotted 
$\langle [X(t)]^{2}\rangle $ as obtained by numerical integration of 
Eq.\ (\ref{ecua2}) 
for (a) $k_{b} T=0.2$, (b) $k_{b} T=0.4$, 
(c) $k_{b} T=0.6$ and (d) $k_{b} T=0.8$. Overimposed to these lines 
the linear regression of the numerical results 
for $t\ge30$ is shown (long-dashed line). 
The solid line is the theoretical prediction $\langle [X(t)]^{2}\rangle $ 
from  Eq.\ (\ref{corrx}); this line practically overlaps with the linear 
regression in Fig.\ a, b, and c. The first-order result 
$\langle [X(t)]^{2}\rangle $ from Eq.\ (\ref{ecua14}) is shown as a dot-dash 
line.}  
\label{graph1}
\end{center}
\end{figure}

Figures \ref{graph1}(a)-(d) show a comparison of our numerical results with the 
analytical predictions, Eqs.\ (\ref{ecua14}) and (\ref{corrx}), 
for different values of $k_{b} T$. We see that there is an excellent 
agreement between theory and numerics except for the highest value of
$k_bT$ [Fig.\ \ref{graph1}(d)]. We have checked that this disagreement 
arises from the way we compute the kink center: For such large values of
the noise, points where $\phi(x,t)=\pi$ are found all over the system, 
irrespective of their distance to the kink center (we note, however, that
the temperature was not high enough to create new kink-antikink pairs).
Those points contribute to the center position through our averaging 
procedure, and in fact their contribution can be shown to be additive, 
i.e., it amounts to move the whole curve 
$\langle [X(t)]^{2}\rangle $ upwards. This is indeed what occurs in 
Fig.\ \ref{graph1}(d), and as we will see below the slope is very close
to the predicted one. The same behavior is found for higher temperatures
in so far no new kinks are created (not shown). Interestingly, 
a first conclusion that can be drawn 
from these figures is that already for not so high temperatures, 
$k_bT\ge 0.4$, as time passes the kink behavior becomes more and more
different from the first-order prediction, showing clearly the 
necessity for the second-order correction.  

\begin{figure}
\begin{center}
\hspace{-2.4cm}
\epsfig{file=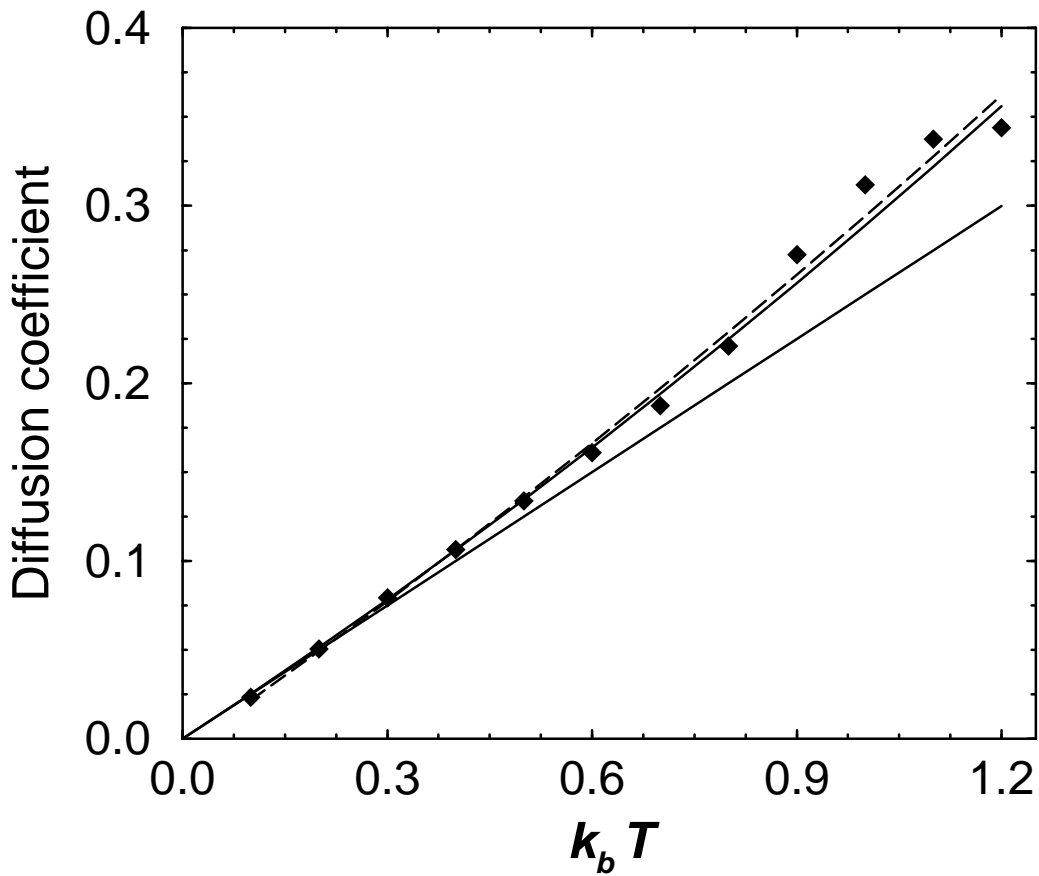, width=5.0in, angle=-90}
\ \\[15mm]
\caption{Lower solid line: the function 
$D_{1}$; upper solid line:
$D_{2}$, which represent the first- and second-order results for the kink
diffusion coefficient [see Eqs.\
(\ref{ecua14}) and (\ref{corrx})]. Diamonds
represent the numerical values of 
the kink diffusion coefficient, obtained by numerical integration of Eq.\ (\ref{ecua2})  
with final time 
$t_{f}=200$ (as in Fig.\ \ref{graph1})
and different values of $k_{b} T$. A quadratic regression 
of these numerical values is also plotted (long-dashed line).}
\label{graph2}
\end{center}
\end{figure}

We have calculated the numerical values of the diffusion coefficient for 
several temperatures by taking the slope of $\langle [X(t)]^{2}\rangle $, 
which we obtain from a linear fit of the data for not so early times  
($t \ge 30$) 
to avoid transient effects coming from the adjustment of the kink to the
heat bath. Note also that our prediction for the second-order contribution
was obtained in the large-time limit, so we should not try to fit the 
whole evolution. 
The figures also show those linear regressions. 
Subsequently, in Fig.\ \ref{graph2} we have compared the computed slopes  
with the first and second-order coefficients  
$\displaystyle{D_{1}=\frac{k_{b} T}{4 \alpha}}$, and 
$\displaystyle{D_{2}=\frac{k_{b} T}{4 \alpha} 
\Big\{ 1+ \Big(\frac{3}{32} + \frac{3}{128} 
\sigma^{2}\Big) k_{b} T \Big \}}$ [see. Eqs.\ (\ref{ecua14}) and 
(\ref{corrx})].  The comparison is once again very good, and points out
very clearly that for values of $k_bT$ as low as 0.3, the first-order 
prediction begins to deviate from the diffusion constant measured in
the simulations. In 
addition, the quadratic fit to the simulation results, shown as a 
long-dashed line in Fig.\ \ref{graph2}, practically coincides with the 
second-order prediction in the whole studied range.

\begin{figure}
\begin{center}
\hspace{-2.4cm}
\epsfig{file=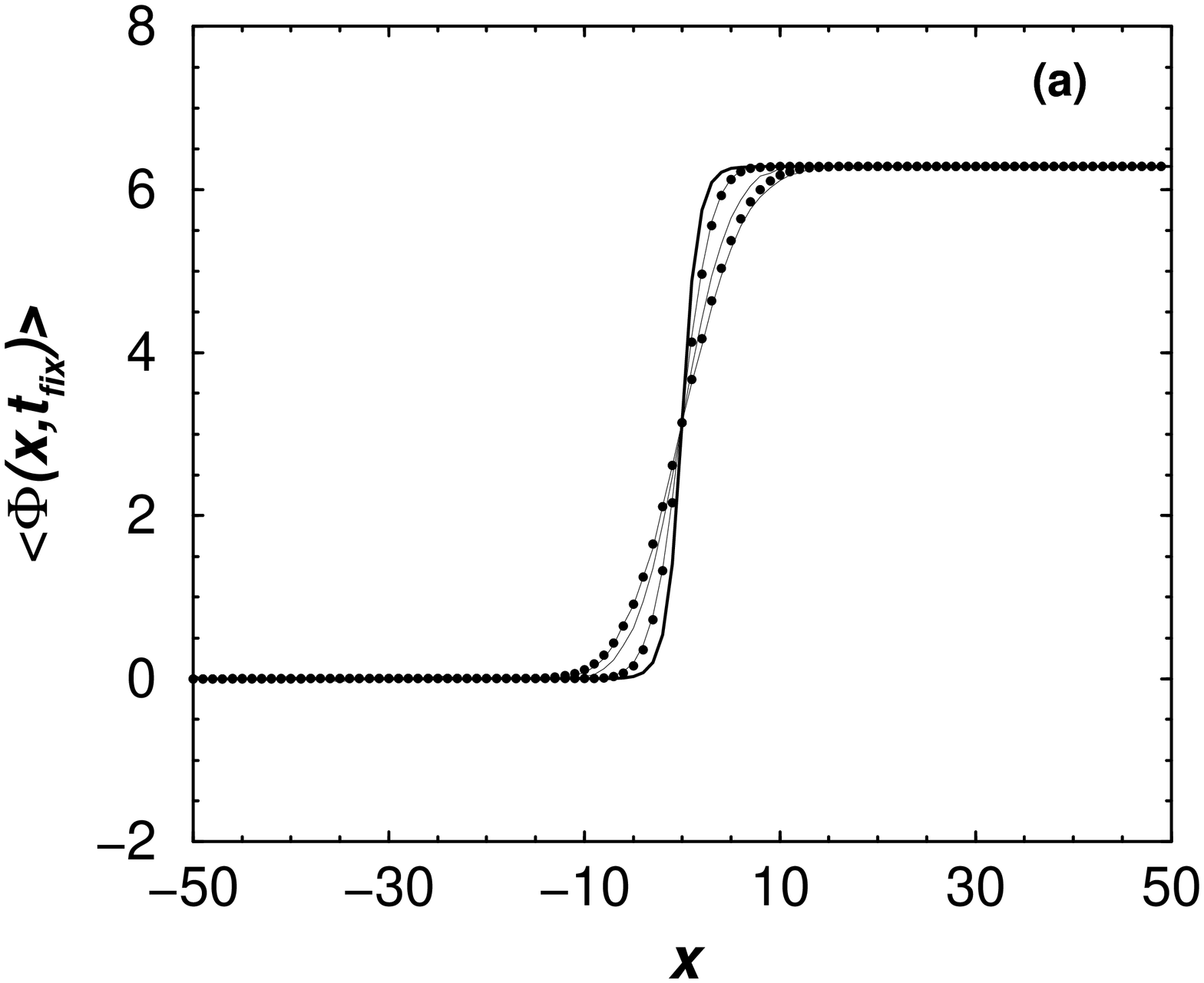, width=2.7in, angle=-90}
\epsfig{file=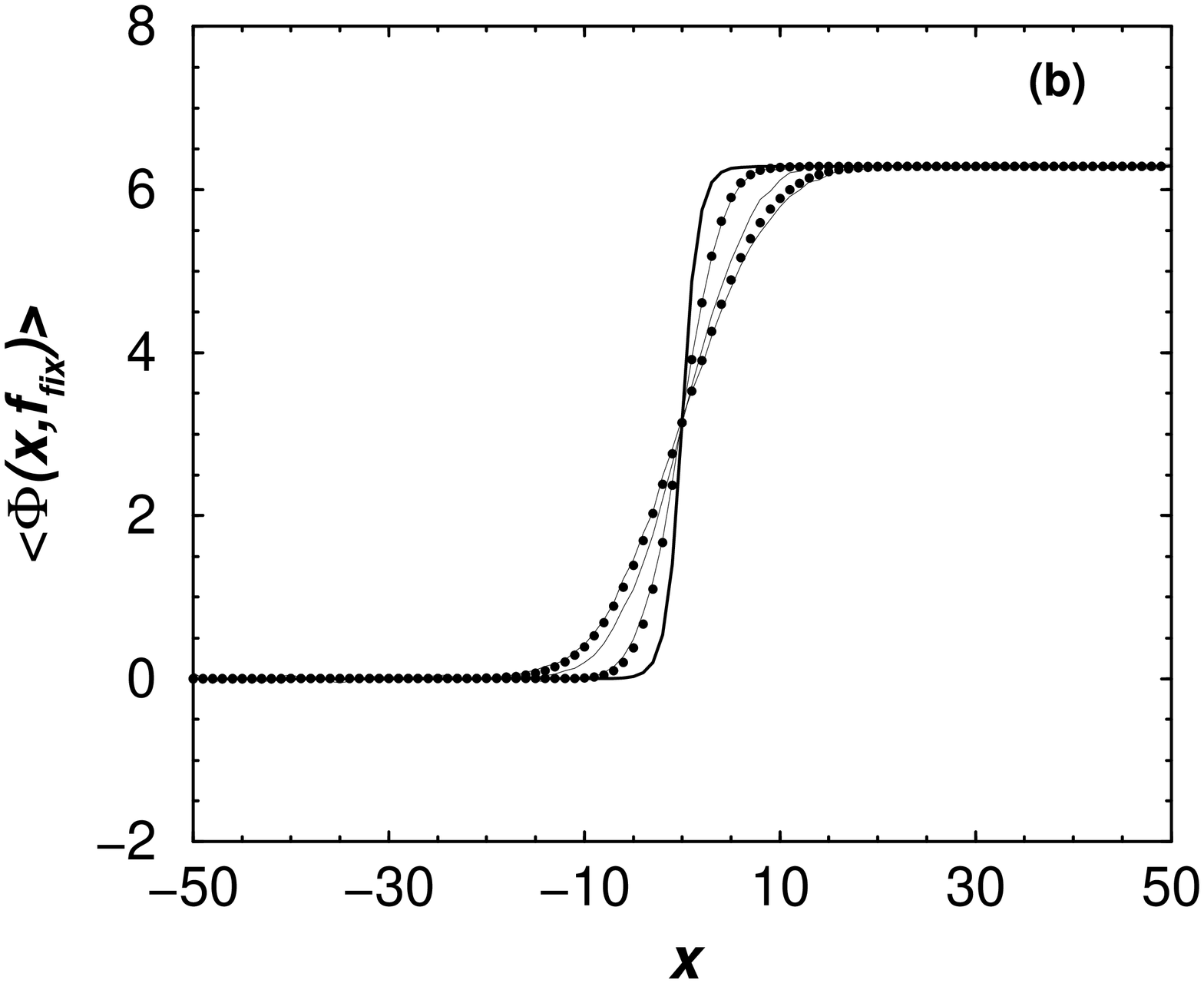, width=2.7in, angle=-90}
\ \\[15mm]
\caption{Solid lines: Snapshots of the evolution of 
$\langle \phi(x,t)\rangle $, obtained from numerical 
simulations of the partial differential equation, for fixed times 
40, 120 and 200, respectively. The initial kink (unperturbed, at rest, 
is also included for comparison). The width of $\langle \phi \rangle $ 
increases as
time progresses.
The overimposed points have been computed numerically from 
the integral (\ref{inte}). Plots correspond to $k_{b} T=0.4$
(a), and 0.8 (b); the width of $\langle \phi \rangle $ 
is seen to increase also with temperature.}
\label{graph3}
\end{center}
\end{figure}

As a final verification of our results, in Fig.\ \ref{graph3}
we have plotted the mean value $\langle \phi(x,t) \rangle$ of the wave 
function at three different
times along its evolution, both as obtained from the 
numerical simulation of the partial differential equation and 
{}from the numerical evaluation of Eq.\ (\ref{inte}). 
The perfect agreement between these expressions provides us with a hint 
to derive an approximate analytical estimate of the evolution 
of $\langle \phi \rangle$ 
{}from the integral (\ref{inte}). 
{}From Fig.\ \ref{graph3} one immediately concludes that the width of $\langle \phi \rangle$  
increases with temperature and time. Let us 
define the width of $\langle \phi \rangle$ by 
\begin{equation}
\displaystyle {
L(t)=\sqrt{\frac{\int_{-\infty}^{+\infty} x^{2} \langle [\phi_x(x,t)]^{2}\rangle  dx}
{\int_{-\infty}^{+\infty} \langle [\phi_x(x,t)]^{2}\rangle  dx}}.
}
\label{anwidth}
\end{equation}

With this definition, we can now calculate $\langle [\phi_x(x,t)]^2\rangle $ 
by using the distribution function of $X_{1}(t)$; this procedure yields
\begin{equation}
\displaystyle {
L(t) \approx 
\sqrt{L_{0}^{2}+\langle [X_{1}(t)]^{2}\rangle }}, 
\label{eqL}
\end{equation}
where 
${\displaystyle L_{0}^{2}=\frac{\int_{-\infty}^{+\infty} dx \,\ 
[x^{2}/\rm cosh^{2}(x)]}
{\int_{-\infty}^{+\infty} dx \,\ [1/\rm cosh^{2}(x)]}=0.8225}$. 

It is important to note that, of
course, we could define $L(t)$ using $\langle \phi_x(x,t)\rangle $ 
instead of $\langle [\phi_x(x,t)]^{2}\rangle $ in the above expression, 
or equivalently another quantity which is localized around the kink 
center. 
However, as all possible (and sensible)
definitions of $L(t)$ give more or less the same results, 
the difference between them becomes a constant factor 
when $\langle [X_{1}(t)]^{2}\rangle $ increases above 
$L_{0}^{2}$ (for example, for large enough $t$). 
So, we expect that the ratio
\begin{equation}
\label{ratio}
\frac{L(t)}{L(t_{fix})} \to \sqrt{\frac{t}{t_{fix}}}, 
\end{equation}
for large enough $t$ and $t_{fix}$. 

\begin{figure}
\begin{center}\hspace{-2.4cm}
\epsfig{file=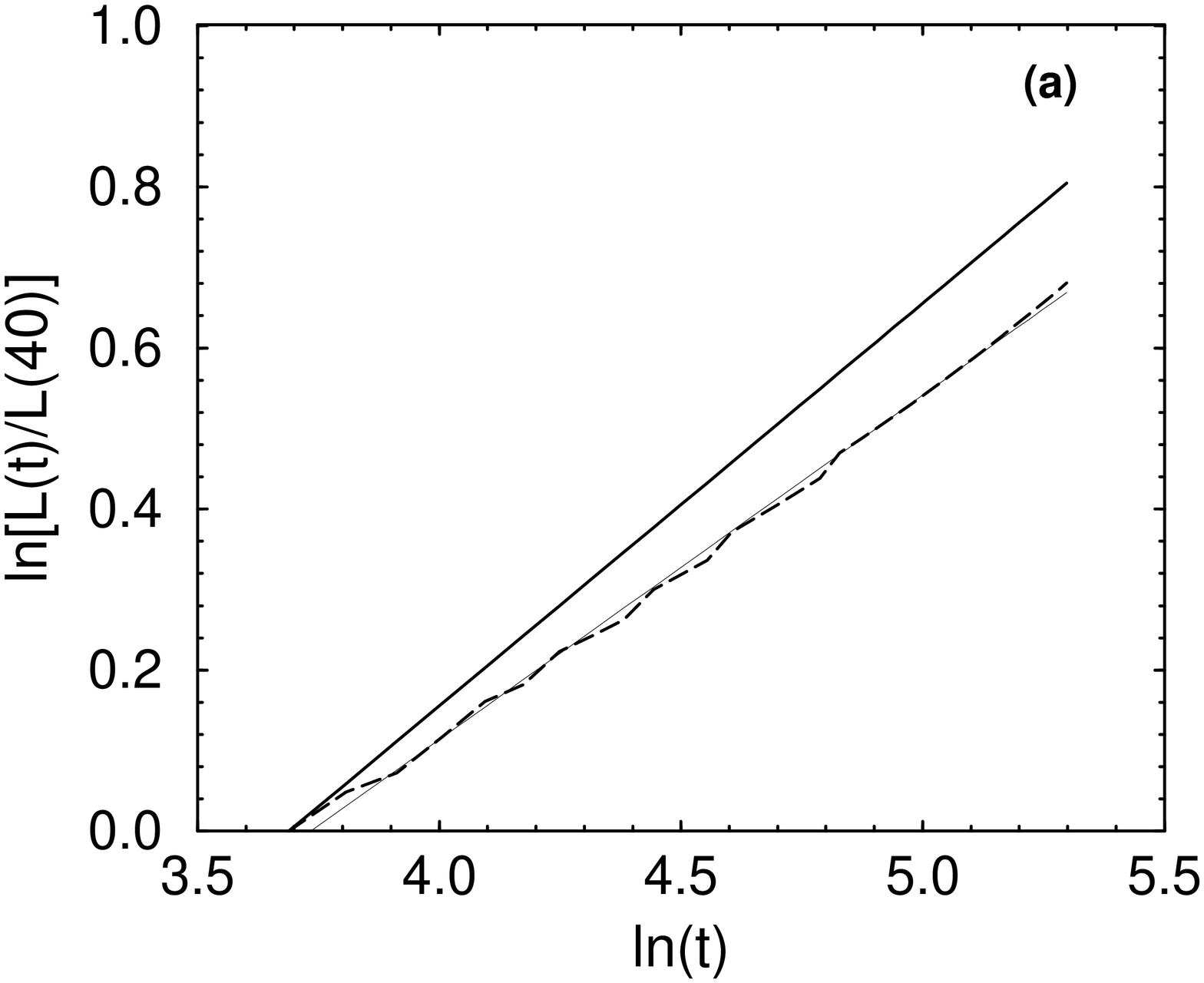, width=2.7in, angle=-90}
\epsfig{file=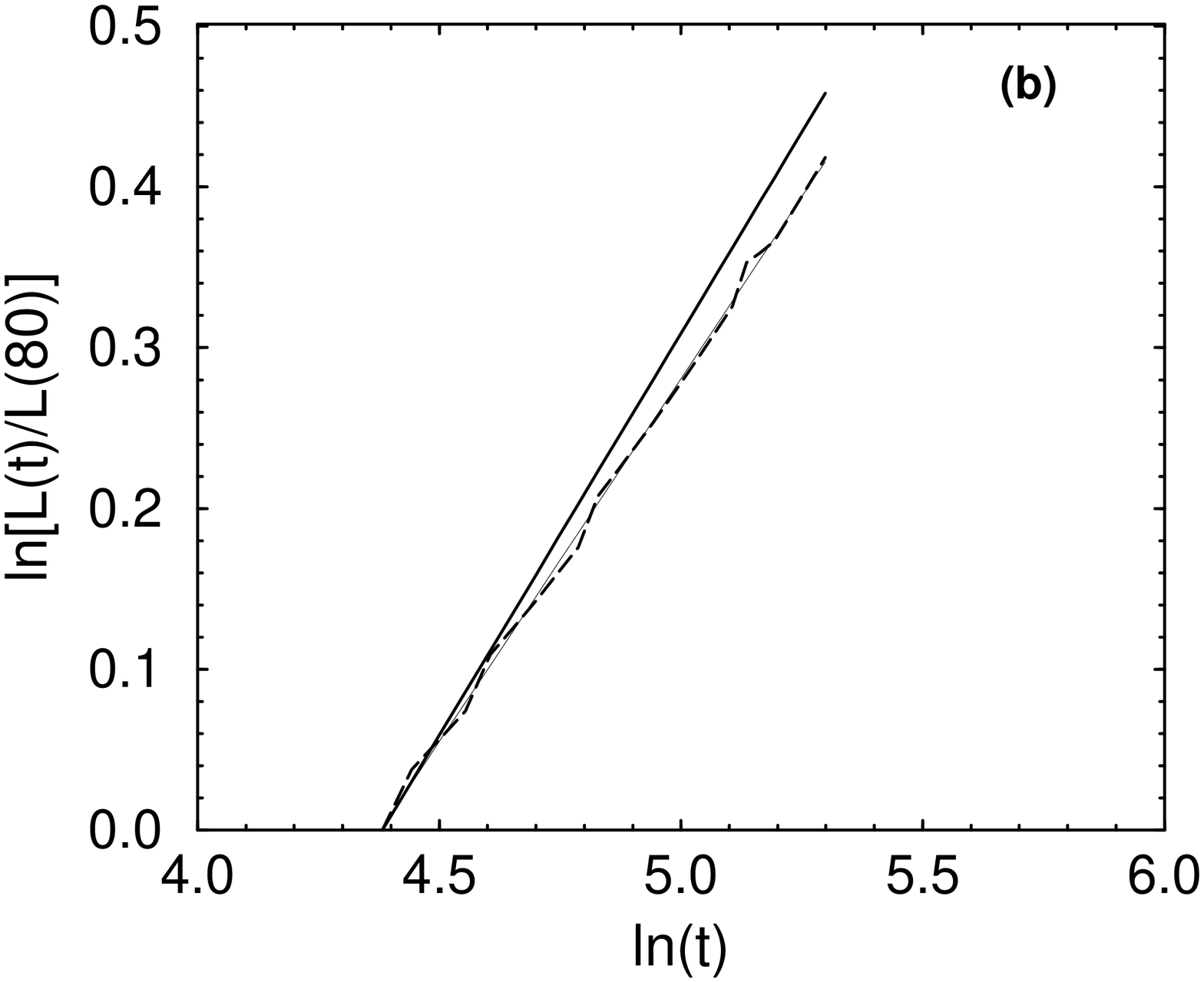, width=2.7in, angle=-90}
\ \\[15mm]
\caption{Solid lines: Analytical values of  
$\rm ln[L(t)/L(t_{fix})]$ for $t_{fix}=40$ (a) and $t_{fix}=80$ (b). 
In both cases $\alpha=1$ and $k_{b} T=0.6$.  
Long-dashed lines: numerical values, calculated from (\ref{inte}). 
The solid lines over the long-dashed ones correspond to 
the linear regression of the
numerical points.}
\label{graph4}
\end{center}
\end{figure}

Figure \ref{graph4} shows a comparison of this prediction 
with the numerical evaluation of the width of $\langle \phi \rangle $ from Eq.\ (\ref{inte}).  
{}From these plots we see that the  
broadening of $\langle \phi \rangle $ behaves indeed as $\sqrt{t}$:
We can compare the analytical slope equal to $0.5$ with 
the numerical ones equal to $0.4276$ and $0.4517$ for plots (a) and 
(b) respectively. The slope in (b) is closer to the analytical value 
due to the fact that $t_{fix}$ is larger than in case (a),
which agrees with the above considerations. 
 
\section{Discussion and conclusions}

As we have seen in the previous section, our second-order theoretical 
predictions constitute a very accurate description of the kink dynamics
for a wide range of temperatures, up to a value of $k_bT\simeq 1$. 
In fact, the range of validity of the analytical results might be
somewhat higher, provided a better way to estimate the kink center from the 
numerical simulations could be devised. In any event, the occurrence of 
$\pi$-crossings far away from the kink center for values around
$k_bT\simeq 1$ indicates that further increments of the temperature
would undoubtedly produce kink-antikink pairs, thus invalidating our 
collective coordinate approach which necessarily relies on the identification
of the individual kink propagation. We note that this value is a little 
over 10\% of the kink rest mass ($M_0=8$ in our units); in this respect, a similar 
result was obtained in \cite{jacek} for the overdamped $\phi^4$ model
by means of a similar perturbative approach (with the caveat that the numerical 
data presented in \cite{jacek} only allow one to guess what is the range of 
validity of their results). 

It is interesting to pursue further the comparison of the results for 
the sG and $\phi^4$ cases. In our calculations for sG, we have found 
that the second-order correction is clearly smaller (albeit relevant)
than the first-order one. The structure of the perturbative calculation 
allows to identify the origin of that correction: It comes from the 
interaction of the phonons (described by the functions $A_k$) with 
the kink. Now, in the $\phi^4$ case, the situation is quite different: 
Indeed, the second-order correction is much {\em larger} than the one 
we find here, and the reason is the so-called internal mode, 
present for $\phi^4$ kinks and absent in the sG case. The coupling 
between this internal mode (which has been shown to act as a reservoir
of energy available for exchange with the kink translation mode 
\cite{Campbell}) and the kink motion can be shown, by a careful examination
of the calculation in \cite{jacek}, to be responsible 
for most of the second-order correction, while the phonons produce a 
second-order term comparable to the one we have found. We thus see that,
while the range of validity of the analytical approach is in principle 
the same in both cases, the physics is certainly different, and in fact
the question arises as to the validity of this kind of perturbative 
calculation for the $\phi^4$ problem in view of the large contribution
of the internal mode. This is an interesting question that deserves 
further analytical and numerical work. 

Coming back to our results for the sG kink, the fact that the second-order
correction is smaller than the first-order term makes us confident that 
our expansion is likely to be free of problems coming from secular terms.
This belief is reinforced by the result that, up to the validity range 
discussed above and limited by kink-antikink creation phenomena, the 
second-order result describes very accurately the kink behavior, which 
deviates very little from the predicted diffusive motion. It is then 
reasonable to expect higher-order contributions (whose calculation is 
extremely cumbersome, but feasible in principle) to be negligible,
thus yielding our theoretical result as the final one for the kink 
diffusion in the overdamped sG problem. In this context, it is also important to 
realize that Eqs.\ (\ref{ecua8})-(\ref{ecua9}), which are only first-order, 
can also be obtained  
following the McLaughlin and Scott procedure \cite{McL} (see also 
\cite{siam}). However, the advantages  
of the perturbative scheme we have used are, on one hand, 
that we were able to obtain the next order in the expansion, 
and on the other hand, we demonstrated that the second order originates in  
the interaction between phonon and translational modes of the sG  
kink. 

\begin{figure}
\begin{center}
\hspace{-2.4cm}
\epsfig{file=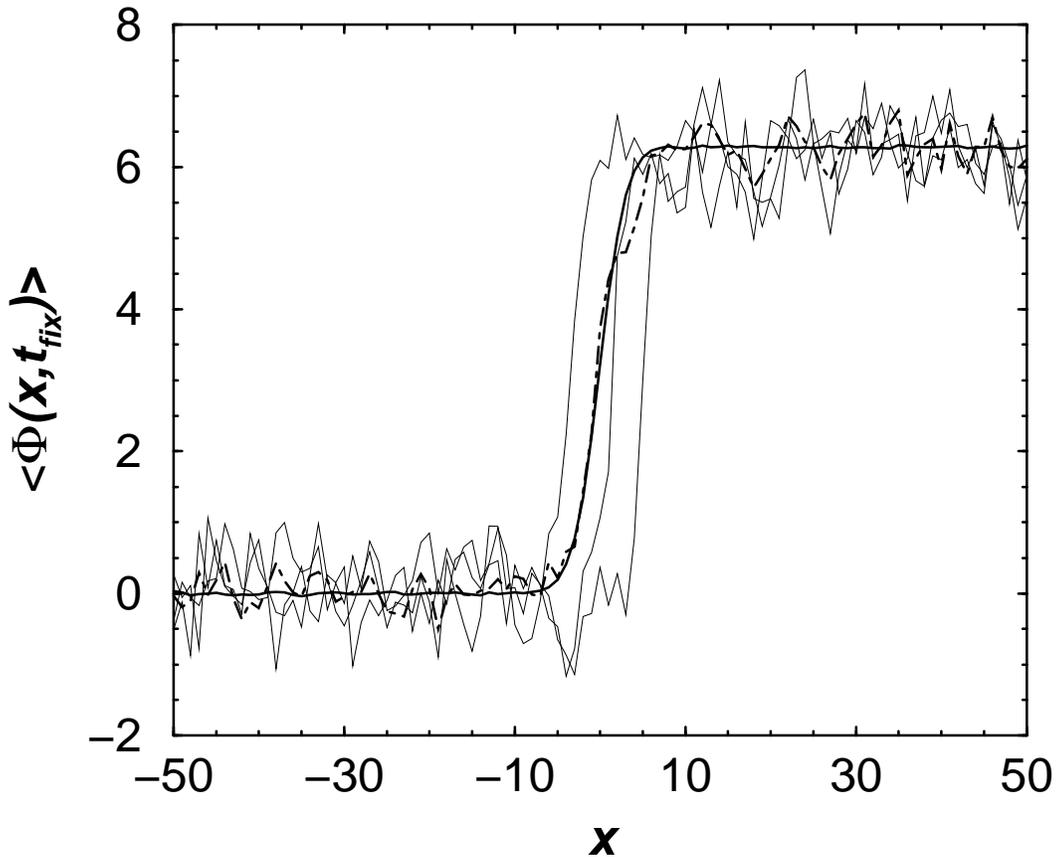, width=5.0in, angle=-90}
\ \\[15mm]
\caption{Average of the wave function for $k_{b}T=0.4$ and $t=200$ 
obtained from 1000 realizations (wider solid line), compared to 
the average of only 5 realizations (dot-dashed line). Also represented 
are 3 of these individual realizations. Note the different slope
and width of the average values as compared to individual realizations.
}
\label{graph5}
\end{center}
\end{figure}

A final remark on our results relates to the mean value of the wave function 
$\langle \phi(x,t)\rangle $ as a function of $t$, 
that must {\it{not}} be interpreted as the shape 
of the kink; in contrast to the interpretation in \cite{jacek}. We first note that the width of the kink 
cannot increase from its value when unperturbed at rest; the sG equation,
being Lorentz invariant, implies that the kink width diminishes when in 
motion, and therefore an increasing of the width would be very difficult to 
understand on physical grounds. Indeed that is not the case. The 
broadening of the mean wave function comes in fact from the dispersion
of the individual realizations, as is immediately seen from Fig.\ 
\ref{graph5}. As may be seen all individual realizations show a width comparable 
to the initial kink width, which agrees with our physical intuition. The
observed 
$\sqrt{t}$ behavior, discussed at the end of the preceding section, 
is then evidently related to the fact 
that the variance of the kink position has that behavior too. The correct
interpretation of the width of $\langle \phi(x,t)\rangle $ is that it represents the
area in which the kink can be located as its diffusive motion
progresses. A similar result has been found for multiplicative noise 
in \cite{armero} (see also \cite{GSbook} and references therein). 

To conclude, we want to stress that our main result is the quadratic 
dependence of the diffusion constant on the temperature, stemming from the kink-phonon 
interactions. This has been verified numerically to a high degree of 
accuracy.  We have carried out 
standard Langevin dynamics simulations following a well grounded 
procedure, the Heun method, as far as statistical properties are 
concerned \cite{maxra}. We can thus be sure that what we are dealing with is 
indeed the dynamics of a sG kink at finite temperature. Therefore, our
analytical calculations and our numerical simulations establish firmly 
the quadratic dependence of the kink diffusion constant on the temperature for the first
time. Now the question remains as to the behavior of {\em underdamped}
sG kinks. Preliminary
calculations \cite{us} seem to indicate that for underdamped sG kinks the 
second-order correction is of the same order as that found here,
which would support the applicability of the previous calculations at least
for small temperatures and not too small damping.  To date, no detailed comparison with 
numerical simulations has ever been done to check the importance of 
the second-order correction. On the other hand, it would be interesting to compare 
the results of our approach with the theoretical analysis and experiments in 
\cite{cast}. Such comparison would provide much insight into the importance of second 
and higher-order corrections in actual physical systems. 
Work along these lines is in progress \cite{us} 

\section*{Acknowledgement}
We are grateful to Esteban Moro, Grant Lythe, and Jos\'e Mar\'\i a Sancho for
discussions. 
Work at GISC (Legan\'es) has been supported by CICyT (Spain) grant MAT95-0325
and DGES (Spain) grant PB96-0119. Travel between Bayreuth and Madrid is 
supported by ``Acciones Integradas Hispano-Alemanas'', a joint program of
DAAD (Az.\ 314-AI) and DGES. This research is part of a project supported 
by NATO grant CRG 971090.

\bigskip

{\em Note added in proof:} After acceptance of this paper, we have 
implemented an improved algorithm for detecting the kink center in 
our code. With this new procedure, no spurious contributions (see 
discussion below Fig.\ \ref{graph1}) to the
variance appear: Specifically, Fig.\ \ref{graph1}(d) is largely 
improved, and the numerical results overlap the theoretical prediction,
thus confirming the interpretation we have provided of the discrepancy. 
A detailed report will be given in \cite{us}.

\section*{Appendix I}

One class of solutions of (\ref{ecua2}) [with $\epsilon=0$] is represented by 
a static kink 
\begin{equation}
\phi_{0}(x,t)= 4 \,\ \mbox{arctan}[\exp(x)].
\label{ap1}
\end{equation}

The perturbations over this equation may be treated by assuming that the solution of 
(\ref{ecua2}) [with $\epsilon=0$] has the form 
\begin{equation}
\phi(x,t)=\phi_{0}(x) + \psi(x,t), \,\ \,\ \psi(x,t) \ll \phi_{0}(x).
\label{ap2}
\end{equation}
If we substitute Eq. (\ref{ap2}) in (\ref{ecua2}) [with $\epsilon=0$] and linearize 
around $\phi_{0}(x)$, we obtain the following equation for $\psi(x,t)$ 
\begin{equation}
\alpha \psi_{t} = \psi_{xx} - \Big[1-\frac{2}{\cosh^{2}(x)}\Big] \psi.
\label{ap3}
\end{equation}
Then, the solution of (\ref{ap3}) may be written as  
$\displaystyle {\psi(x,t) = f_{k}(x) 
\exp\Big(-\frac{\omega_{k}^{2} \,\ t}{\alpha}\Big)}$, where $f_{k}(x)$ satisfies 
the eigenvalue problem given by
\begin{equation}
-\frac{\partial^{2} f_{k}}{\partial x^{2}} + \Big[1-\frac{2}{\cosh^{2}(x)}\Big] 
f_{k} = 
\omega_{k}^{2} f_{k}.
\label{ap4}
\end{equation}
This equation admits the following eigenfunctions with their respective 
eigenvalues 

\begin{eqnarray}
f_{T}(x) = \frac{2}{\cosh(x)}, \,\ \omega_{T}^{2} & = & 0, \\
f_{k}(x) = \frac{\exp(i k x) \,\ [k + i \mbox{tanh}(x)]}{\sqrt{2 \pi} \,\ \omega_{k}}, \,\ 
\omega_{k}^{2} & = & 1 + k^{2}.
\label{ap5}
\end{eqnarray}

Notice, that $f_{T}(x)$ and $f_{k}(x)$ form a complete set of functions with 
the orthogonality relations 
\begin{mathletters}
\begin{eqnarray}
\int_{-\infty}^{+\infty} f_{T}^{2}(x) \,\ dx & = & 8, \quad 
\int_{-\infty}^{+\infty} f_{T}(x) f_{k}(x) \,\ dx = 0, \\
\int_{-\infty}^{+\infty} f_{k}(x) f_{k'}^{*}(x) \,\ dx & = &\delta(k-k').  
\end{eqnarray}
\label{ap6}
\end{mathletters}


\end{document}